\documentclass[aps,prb,twocolumn,groupedaddress]{revtex4-1}
\usepackage{bm}
\usepackage{graphicx}
\usepackage[usenames]{color}
\usepackage{amsmath,amsfonts,amssymb}
\usepackage{hyperref}
\usepackage{ulem}
\usepackage{xcolor}

\newcommand{\ket}[1]{|{#1}\rangle}

\newcommand{\pa}{\partial}
\newcommand{\Tr}{\text{Tr\,}}
\allowdisplaybreaks[4]

\begin{document}
\title{Adiabatic Heuristic Principle on a Torus and Generalized Streda Formula}

\author{Koji Kudo$^1$}
\author{Yasuhiro Hatsugai$^{1,2}$}
\affiliation{
$^1$Graduate School of Pure and Applied Sciences, University of Tsukuba, 
Tsukuba, Ibaraki 305-8571, Japan\\
$^2$Department of Physics,
University of Tsukuba, Tsukuba, Ibaraki 305-8571, Japan
}

\date{\today}

\begin{abstract}
 Although the adiabatic heuristic argument of the fractional quantum Hall 
 states has been successful, continuous modification of the flux/statistics of 
 anyons is strictly prohibited due to algebraic constrains of the braid group 
 on a torus. We have numerically shown that the adiabatic heuristic
 principle for anyons
 is still valid even though the Hamiltonians cannot be modified continuously.
 The Chern number of the ground state multiplet is the adiabatic 
 invariant, while the number of the 
 topological degeneracy behaves wildly. A generalized Streda formula is 
 proposed that explains the degeneracy pattern. Nambu-Goldston modes associated
 with the anyon superconductivity are also suggested numerically.
\end{abstract}

\maketitle

\section{Introduction}
Over the past decade, topology has been coming to the fore in modern 
condensed matter physics. The quantum Hall (QH) effect~\cite{
Klitzing_IQH_PRL80,Tsui_FQH_PRL82} is a prime example of topologically 
non-trivial phases, where the quantized Hall conductance is given by the Chern 
number~\cite{Thouless_TKNN_PRL82,Kohmoto_Chern_Ann85,Niu_NTW_PRB85,
Berry_BP_PRC84}. Topological concepts 
enrich material phases beyond the Ginzburg-Landau theory. The fractional QH 
(FQH) state~\cite{Laughlin_FQH_PRL83} is a typical example of the quantum 
liquid with the topological order~\cite{Wen_TO_PRB89}. It hosts
fractionalized excitations that carry fractional charges and fractional 
statistics~\cite{Arovas_FS_PRL84,Haldane_boson_PRL83,Halperin_FS_PRL84}, which 
is the hallmark of the topologically ordered phases~\cite{Wen_TO_AP95}. 
The topological degeneracy is closely related to these 
fractionalizations~\cite{Einarsson_Braid_PRL90,Wen_TO_PRB89,Oshikawa_TO_PRL06,
Sato_TO_PRL06}.
Some of the non-Abelian topological order can be used for a possible quantum 
computation~\cite{Willett_5/2_PRL87,Moore_MR_NPB91,Read_RRstate_PRB99,
Kitaev_TQC_Ann03,Nayak_TQC_PMP08}.

Point particles in two-dimension can be charge-flux composites associated with
a singular gauge transformation~\cite{Wilczek_anyon2_PRL82}. In 
relation to the composite fermion picture~\cite{Jain_CFT_PRL89,jain_2007}, the
flux-attachment has been quite successful to describe the FQH effect;
the FQH effect at the filling factor $\nu=p/(2mp\pm1)$ with $p$ and $m$ 
integers can be understood as the $\nu=p$ IQH effect of the composite fermions.
This concept is further developed to the ``adiabatic heuristic 
principle''~\cite{Greiter_AH_NPB90,Greiter_AH_NPB92}. It states that both 
states are adiabatically connected through intermediate systems of anyons.
This characterization of the QH states 
based on the adiabatic deformation is a typical example of the topological 
classification as is widely applied to the recent studies of topological 
phases.

We note that a careful setup is required to carry 
the program of this adiabatic heuristic principle for concrete systems. 
The statistical phase $\theta$ of anyons is governed by a representation of the
fundamental group of the many-particle configuration space 
(braid group)~\cite{Wu_Braid_PRL84}. Therefore, the world 
lines of the system needs to satisfy the braid group constraint.

As for topological phenomena, the geometry of the system is crucially 
important. With boundaries, low energy modes appear as edge states even for
gapped systems. 
Thus, for the demonstration of the adiabatic heuristic principle, the 
torus geometry without any boundaries is favorable~\cite{torus}. 
However, an algebraic constraint of the braid group on a 
torus~\cite{Birman_Braid_M69,Einarsson_Braid_PRL90,Wen_anyon_PRB90,Hatsugai_anyonTr_PRB91,Einarsson_Braid_MPLB,Li_Braid_B93}
prohibits continuous change of the statistical phase $\theta$. This makes it 
impossible to apply the adiabatic heuristic principle naively.

In this Letter, we show that the adiabatic heuristic principle indeed 
remains valid on a torus. Here,  ``adiabatic'' is used in the sense
that the gap remains open although continuous deformation of the Hamiltonian
is impossible.
The many-body Chern number of the ground state multiplet is also calculated 
numerically, which serves as the adiabatic invariant while their degeneracy
changes wildly. 
We propose a generalized 
Streda formula to characterize the obtained degeneracy pattern in relation to 
the Chern number, which follows from the translational invariance of anyons. 
At the gap closing point, the Chern number changes its sign and the anyon 
superconductivity~\cite{
Laughlin_anyonS_PRL88,Fetter_anyonS_PRB89,CHEN_anyonS_B89} is expected.

\section{Adiabatic heuristic principle and braid group}
Let us here shortly derive the fundamental relation of the adiabatic 
heuristic principle~\cite{Greiter_AH_NPB90,Greiter_AH_NPB92}.
We consider a QH system of $N_a$ particles with the charge $-e$ 
in a uniform magnetic field. According to the adiabatic heuristic principle, 
the QH state is adiabatically deformed by trading the external fluxes 
for the statistical ones of anyons. Since the total flux remains constant
($N_\phi+N_a\theta/\pi=\text{const.}$), one has the relation
$1/\nu+\theta/\pi=\text{const.}$,
where $N_\phi$ is the number of the external flux, $\theta$ is the statistics 
of anyons and $\nu=N_a/N_\phi$.
Assuming that the $\nu=p$ IQH state of fermions ($\theta=\pi$) is included in 
this series, one has $\nu=p/[p(1-\theta/\pi)+1]$.

\begin{figure}[t!]
  \begin{center}  
   \includegraphics[width=\columnwidth]{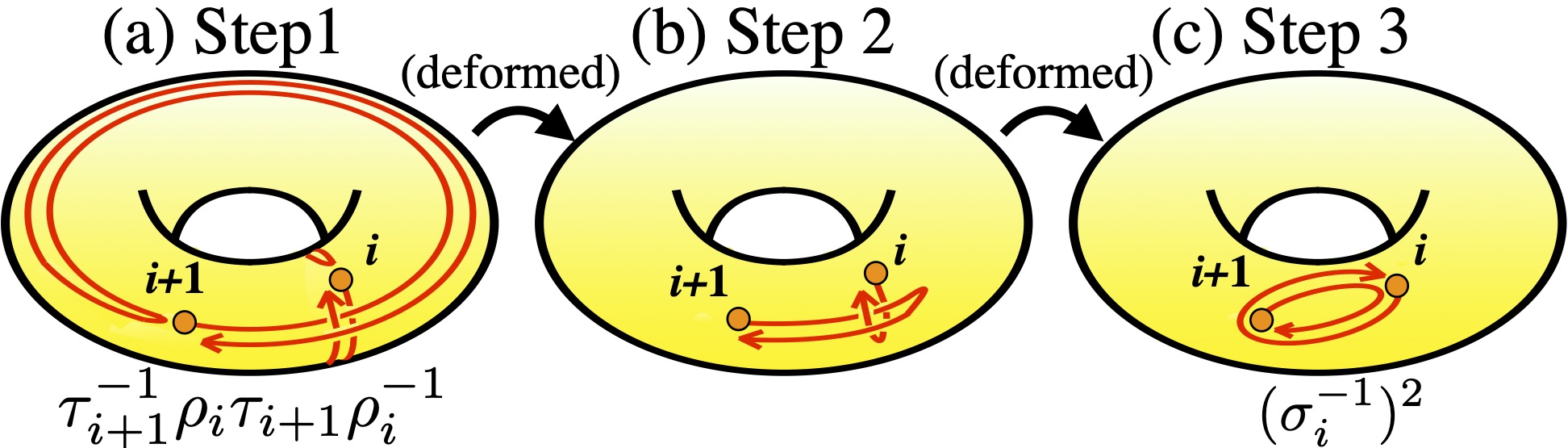}
  \end{center}
 \caption{
 Pictorial proof of Eq.~\eqref{eq:braid}~\cite{Birman_Braid_M69}. (a) The 
 paths of the four moving processes $\rho_i^{-1}$, $\tau_{i+1}$, $\rho_i$, and
 $\tau_{i+1}^{-1}$ are presented in the order from bottom to top. 
 (c) The path in (a) can be deformed into that of $(\sigma_i^{-1})^2$.
 }
 \label{fig:braid}
\end{figure}

From the analysis of the braid group on a 
torus~\cite{Birman_Braid_M69,Einarsson_Braid_PRL90,Wen_anyon_PRB90,
Hatsugai_anyonTr_PRB91,Einarsson_Braid_MPLB,Li_Braid_B93}, 
the relation $1/\nu+\theta/\pi=\text{const.}$ is rederived
(see Appendix~\ref{sec:braid}) with an 
{\it additional} constraint as explained below.
The generators of the braid group on a torus are denoted as $\sigma_i$, 
$\tau_i$ and $\rho_i$, where $\sigma_i$ ($i=1,\cdots,N_a-1$) is a local 
exchange between the $i$th and $i+1$th anyons, and $\tau_i$ and $\rho_i$ 
($i=1,\cdots,N_a$) are global moves of the $i$th anyon along a noncontractible 
loop on the torus in $x$ and $y$ directions.
Now, we take the basis $\ket{\{\bm{r}_k\};w}$ for their expressions, where
$\{\bm{r}_k\}$ is the positions of anyons and $w=1,\cdots,M$ is the extra 
internal index that is necessary to satisfy the braid group constraints on a 
torus as seen below. We assume that anyons are Abelian:
$\sigma_i=e^{i\theta}\bm{1}_M$, where $\bm{1}_M$ is the $M$-dimensional unit 
matrix. 
As shown in Fig.~\ref{fig:braid}~\cite{Birman_Braid_M69},  the generators 
$\sigma_i$, $\tau_i$ and $\rho_i$ need to satisfy
\begin{align}
 \tau_{i+1}^{-1}\rho_{i}\tau_{i+1}\rho_{i}^{-1}
 =(\sigma_i^{-1})^2=e^{-i2\theta}\bm{1}_M.
 \label{eq:braid}
\end{align}
By taking a determinant of Eq.~\eqref{eq:braid}, we have $1=e^{-i2M\theta}$. 
If $\theta/\pi=n/m$ (with $n$, $m$ coprime), the dimension of the representation $M$ needs to be a 
multiple of $m$. This constraint strictly prohibits continuous change of the 
Hamiltonian in the adiabatic heuristic principle.

In this Letter, this puzzle is resolved. Although the Hamiltonian is defined 
only for discrete values of $\theta$ and its dimension behaves wildly, the 
energy gap defined by a dense set of the Hamiltonians is surprisingly smooth 
and finite.
It justifies the adiabatic heuristic principle on a torus. 
We also find the generalized streda formula to explain the wild 
behavior of the degeneracy by the many-body Chern number. 

\section{Model}
We consider the periodic system of anyons in the uniform magnetic field on a 
square lattice with $N_x\times N_y$ sites. The Hamiltonian is 
\begin{align}
 H
 =t\sum_{\langle ij\rangle}
 e^{i\phi_{ij}}e^{i\theta_{ij}}c_i^\dagger c_j\otimes W^{(ij)}
 +V\sum_{\langle ij\rangle}n_in_j\otimes\bm{1}_M,
 \label{eq:ham}
\end{align}
where $n_i=c_i^\dagger c_i$ and $c_i^\dagger$ ($c_i$) is the creation 
(annihilation) operator for a hard-core boson on site $i$. 
The hard-core condition is necessary to ensure consistency with the braid
group.
The Peierls 
phase $e^{i\phi_{ij}}$ is specified by the string gauge~\cite{
Hatsugai_Sum_PRL99} for the external magnetic field. The phase 
$e^{i\theta_{ij}}$ describes 
the statistical phase~\cite{Wen_anyon_PRB90,Hatsugai_anyonTr_PRB91} (see 
the details below). 
$W^{(ij)}$ is an $M$-dimensional matrix~\cite{Hatsugai_anyonTr_PRB91} to 
ensure consistency with Eq.~\eqref{eq:braid}. When $\theta/\pi=n/m$, $M$ is 
fixed to be $m$ as the irreducible representation. We 
set $W^{(ij)}=W_x$ and $W_y$ for $(ij)$ describing a pair of sites across the 
boundary in the $x$ and $y$ directions respectively and otherwise 
$W^{(ij)}=\bm{1}_M$, where 
\begin{align}
 W_x&=\left[                                                  
 \begin{array}{cccc} 
  0 & 1 & \ldots & 0 \\
  \vdots & \vdots & \ddots & \vdots \\
  0 & 0 & \ldots & 1 \\
  e^{i\eta_x} & 0 & \ldots & 0
 \end{array}
 \right],\label{eq:Wx}\\
 W_y&=e^{i\eta_y}\text{diag}[e^{i2\theta},e^{i4\theta},\cdots,e^{i2M\theta}],
 \label{eq:Wy}
\end{align}
and $\vec{\eta}=(\eta_x,\eta_y)$ specifies the twisted boundary conditions.
The Hamiltonian is consistent with Eq.~\eqref{eq:braid}
since we have $W_x^{-1}W_yW_xW_y^{-1}=e^{-i2\theta}\bm{1}_M$ for any 
$\vec{\eta}$.

\begin{figure}[t]
  \begin{center}  
   \includegraphics[width=\columnwidth]{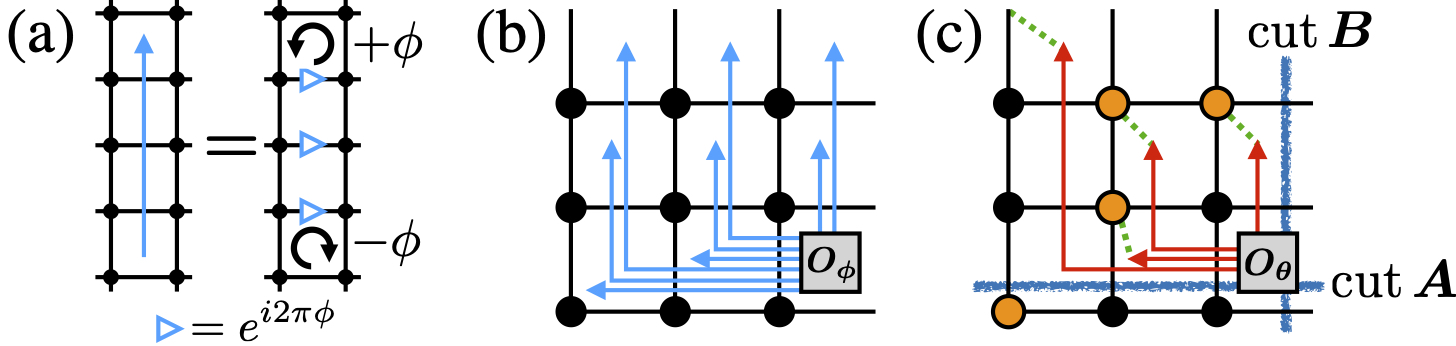}
  \end{center}
 \caption{
 (a) String gauge. (b)(c) Sketches of $3\times3$ square lattice 
 with (b) the string gauge $\phi_{ij}$ and (c) the statistical gauge 
 $\theta_{ij}$. The yellow points in (c) represent sites with anyons.}
 \label{fig:string}
\end{figure}
Let us now give detailed descriptions of how to construct the Hamiltonian
in Eq.~\eqref{eq:ham}. We first mention the string gauge $\phi_{ij}$
briefly.
As shown in Fig.~\ref{fig:string}(a), let us consider a string on sites and 
assigns the Peierls phase
$e^{i2\pi\phi}$ on the links intersected by the string. They clearly describe 
the magnetic fluxes $\phi$ and $-\phi$ at the initial and terminal points of 
the string, respectively. Thus, the string gauge shown in 
Fig.~\ref{fig:string}(b) introduces the flux $\phi\times(1-N_xN_y)$ to the 
plaquette with the origin $O_\phi$ while $\phi$ to the others. 
The gauge convention $\theta_{ij}$ is also described by the strings, see 
Fig.~\ref{fig:string}(c). The strings carry the phase factor $e^{i\theta}$, and
their terminal points are located at plaquettes adjoining anyons.
Besides, the additional rules are given as 
follows~\cite{Hatsugai_anyonTr_PRB91}: (The roles of each rule are explained 
in Ref.~\onlinecite{rules})

(i) If a string sweeps another anyon in the process of hopping, one determines
the phase factor as if the anyon crosses the string.

(ii) When an anyon hops across the cut $B$ from left to right, the phase factor
$e^{i(N_a-1)\theta}$ is given.

(iii) When an anyon hops across a horizontal string, the phase factor not 
$e^{i\theta}$ but $e^{i2\theta}$ is given.

(iv) When an anyon hops across the cut $A$ upward, the phase factor 
$e^{iX\theta}$ is given, where $X$ is the number of other anyons in the same 
$x$-axis position as the hopping anyon.

\begin{figure*}[t!]
   \includegraphics[width=2\columnwidth]{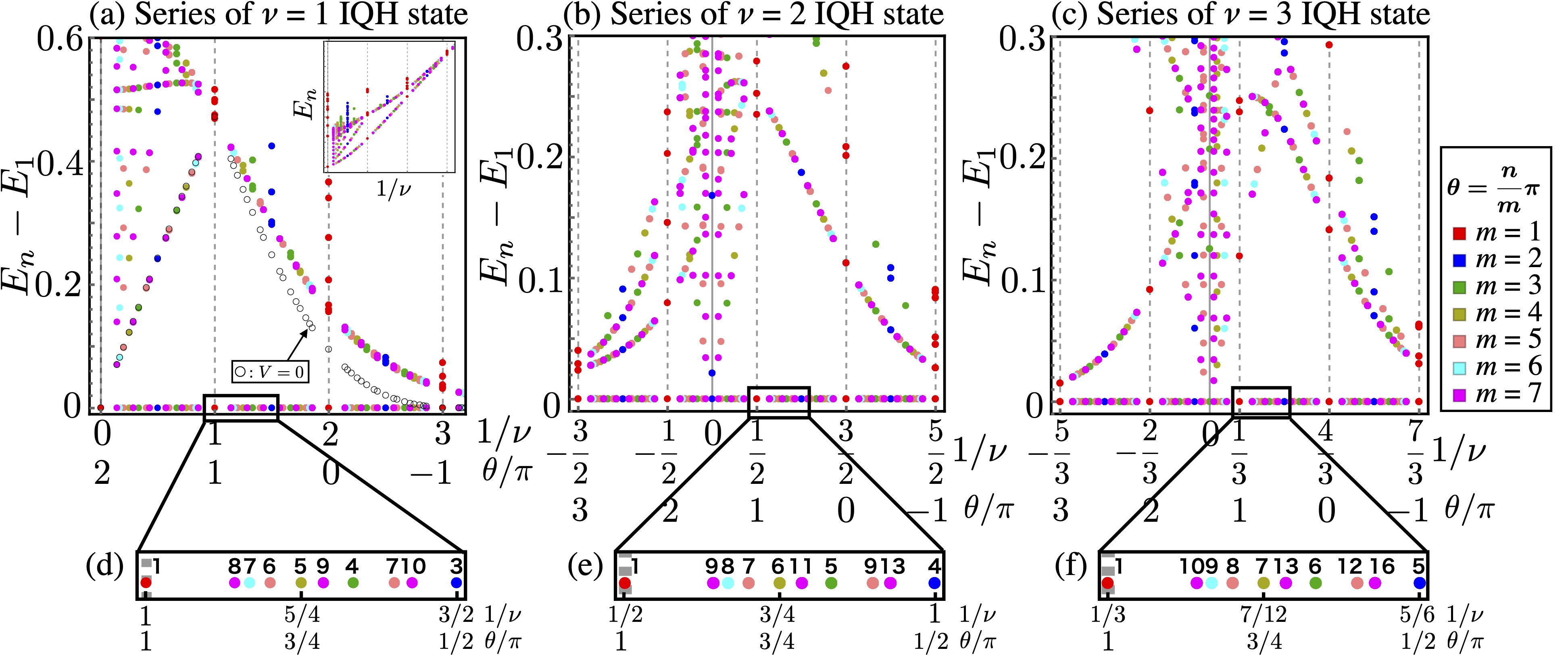}
 \caption{
 (a-c) Energy gaps are shown as functions of $1/\nu$ for $t=-1$ and 
 $V=5$. The system size is $N_x\times N_y=10\times10$. The statistical 
 parameter $\theta$ is determined by $\nu=p/[p(1-\theta/\pi)+1]$ with (a) 
 $p=1$ 
 (b) $p=2$, and (c) $p=3$, respectively. The vertical dashed lines represent 
 $\theta/\pi=\text{integer}$. The anyon number is (a),(b) $N_a=4$ and (c) 
 $N_a=3$. We plot the lowest $N_\text{cut}$ states at each $1/\nu$ in the 
 figures [$N_\text{cut}=40$ for (a) and (b), and $N_\text{cut}=70$ for (c)]. 
 The open black circles in (a) are the gaps $E_{N_D+1}-E_1$ for $V=0$. The 
 inset in (a) is the energy spectrum of the same setting as (a).
 (d-f) Ground state degeneracy $N_D$ is shown.
 }
 \label{fig:ah}
\end{figure*}

Due to Eqs.~\eqref{eq:Wx} and \eqref{eq:Wy}, we also give the following rules:
When an anyon hops across the cut $B$ from left to right, the label is
changed from $w$ to $w-1$, where $w$ is the label of the basis 
$\ket{\{\bm{r}_k\};w}$.
If $w=1$, the phase factor $e^{i\eta_x}$ is also given.
Also, when an anyon hops across the cut $A$ upward, the phase factor 
$e^{i\eta_y}e^{i2w\theta}$ is given. 

In this framework, the representations of the global move operators are given 
as $\tau_j=e^{i\frac{e}{\hbar}\alpha_j}e^{-i2\theta(j-1)}W_x$ and
$\rho_j=e^{i\frac{e}{\hbar}\beta_j}e^{i2\theta(j-1)}W_y$,
where $e^{i\frac{e}{\hbar}\alpha_j}$ and $e^{i\frac{e}{\hbar}\beta_j}$ 
are came from the Peierls phase $\phi_{ij}$ describing the external magnetic 
field. As shown in Appendix~\ref{sec:braid}, these representations are 
consistent with the braid group on a torus.

The above construction of $\theta_{ij}$ introduces 
the magnetic flux $-2\pi\times2\theta N_a$ only to the plaquette with the 
origin $O_\theta$ shown in Fig.~\ref{fig:string}(c)~\cite{
Hatsugai_anyonTr_PRB91}. Since the string gauge $\phi_{ij}$ introduces the flux
$\phi\times(1-N_xN_y)$ to the plaquette with the origin $O_\phi$ while 
$\phi$ to the others as described above, one gets a condition of the uniformity
of the magnetic field as $e^{i2\pi\phi(1-N_xN_y)-i2\theta N_a}=e^{i2\pi\phi}$. 
Since $N_\phi=\phi N_xN_y$, this condition is consistent with the 
relation $1/\nu+\theta/\pi=\text{const.}$

\section{Energy gap}
By the above setup, we numerically diagonalize the Hamiltonians. In the 
following, we set $N_x=N_y=10$, $t=-1$, $V=5$ and $\vec{\eta}=\vec{0}$ unless 
otherwise stated. We assume that the states are degenerate if the energy 
difference is less than $0.001$.

In Figs.~\ref{fig:ah}(a)-(c), we plot the energies of a series that includes 
the $\nu=p$ IQH state ($p=1,2,3$) as a function of $1/\nu$. We show the data 
for $\theta=(n/m)\pi$ with various $m$ and $n$ ($m\leq7$). 
The data points with different colors are eigenvalues of $H$ with the 
different dimensions. Figures~\ref{fig:ah}(a)-(c) show that the gap behaves 
smoothly for a dense set of Hamiltonians.
The energies of the ground state are also smooth, see the inset in 
Fig.~\ref{fig:ah}(a).

\begin{figure}[t!]
  \begin{center}  
   \includegraphics[width=\columnwidth]{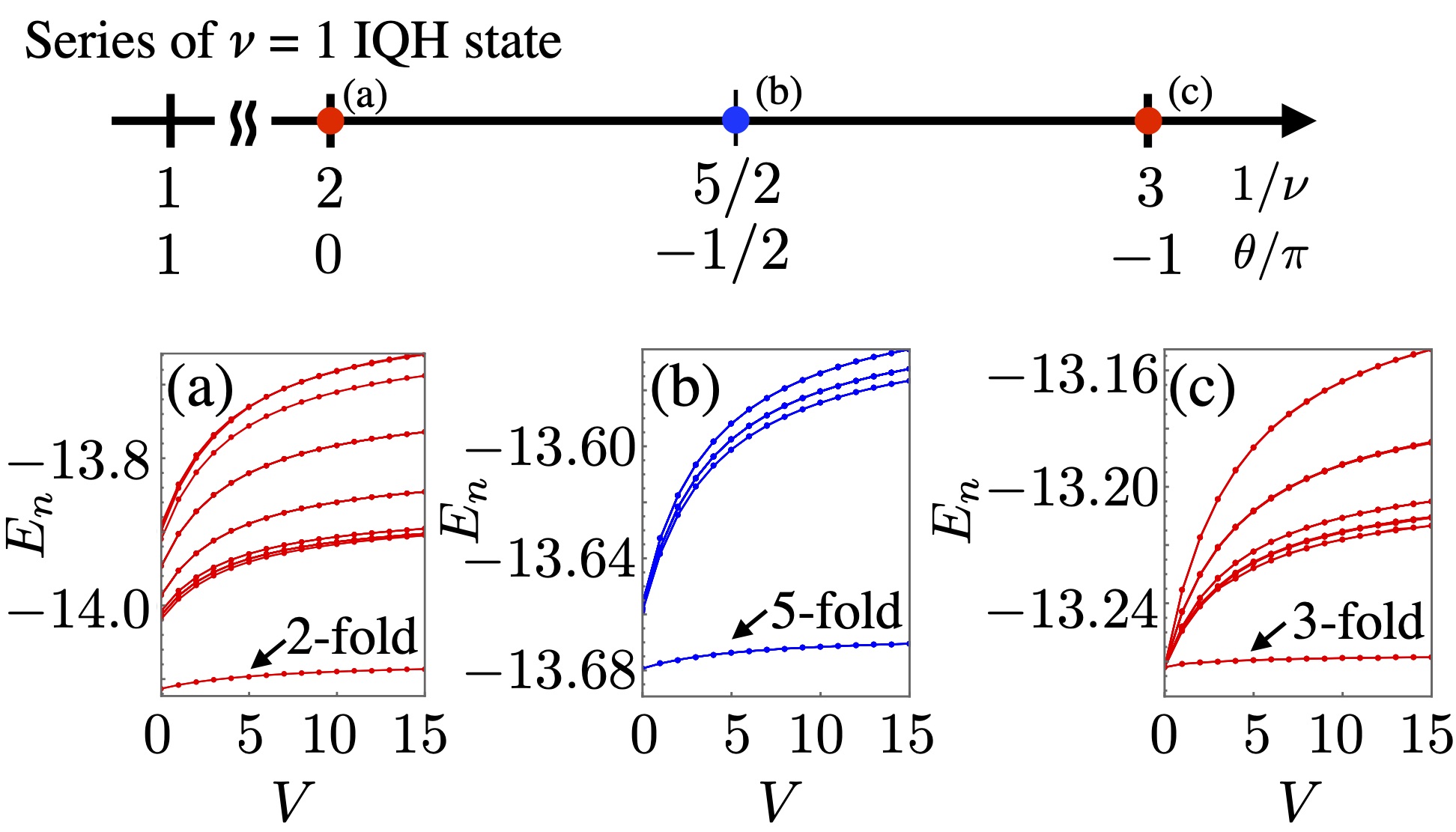}
  \end{center}
 \caption{
 Energy spectra are shown as functions of the interaction $V$ for (a)
 the Boson FQH state at $\nu=1/2$, (b) the anyon FQH state at $\nu=2/5$ and (c)
 the Fermion FQH state at $\nu=1/3$. They are included in a series of the 
 $\nu=1$ IQH state. We set $t=-1$ and $N_x\times N_y=10\times10$. The lowest 
 forty states are shown in each figure.
 }
 \label{fig:t_V}
\end{figure}

Let us first consider a series of the $\nu=1$ IQH state, which includes the 
Laughlin state. We here consider only $0\leq\nu$ since a system of $\nu<0$ is
trivially mapped to that of $0<\nu$. 
In Fig.~\ref{fig:ah}(a), the 3-fold degenerated ground state is obtained at 
$\nu=1/3$ , which is consistent with the lattice analogue of the 
Laughlin state~\cite{Kudo_17}.
This state is adiabatically connected to the $\nu=1$ IQH state. 
Note that, however, the ground state degeneracy $N_D$ changes wildly,
see Fig.~\ref{fig:ah}(d). 
At $1/\nu=0$, the gap closing occurs, which suggests the Nambu-Goldston modes 
associated with the superconductivity of hard-core 
bosons~\cite{Zhang_Boson_B92}. In 
Fig~\ref{fig:ah}(a), the results without the electron-electron interactions
are also shown. 
While the ground states of anyons or bosons are gapped because of their 
hard-core nature, the gap at $\nu=1/3$ vanishes since the system reduces to 
the partially filled lowest Landau band of free fermions. It implies that the 
interaction is crucially important only for the FQH states of fermions. 
In Fig.~\ref{fig:t_V}, the energy spectra as functions of the interaction $V$
are shown. The FQH states remain gapped with the same topological degeneracy
for a wide range of $V$ apart from the point $V=0$ in Fig.~\ref{fig:t_V}(c). 
Inclusion of the finite interaction $V$ induces the gap at this point, which is
consistent with the gapped Laughlin state.
Although the discussion of the thermodynamic limit is an open question, our
adiabatic heuristic argument for the fixed system size includes important
scientific information. 

As for the other series in Figs.~\ref{fig:ah}(b) and (c), one can also see 
that the gaps remain open
for each region $0<1/\nu$ and $1/\nu<0$ although their 
topological degeneracy changes irregularly 
[see Figs.~\ref{fig:ah}(e) and (f)].
The excitation gap 
closes at $1/\nu=0$ in both figures, which is consistent with the emergence of 
the anyon superconductivity~\cite{
Laughlin_anyonS_PRL88,Fetter_anyonS_PRB89,CHEN_anyonS_B89} of
$\theta=(3/2)\pi$ and $\theta=(4/3)\pi$, respectively. 

The numerical results in Figs.~\ref{fig:ah} suggest that the adiabatic
heuristic principle remains valid
for a series that includes $\nu=p$ IQH state for general integer 
$p$. It also suggests the 
realization of the anyon superconductivity of $\theta=(1+1/p)\pi$ by trading 
all the external magnetic flux for the statistical one. 

Since $1/\nu\propto\phi$, where $\phi=N_\phi/(N_xN_y)$ is the number of the 
flux per plaquette, this unusual but adiabatic behavior, in a
sense that the gap remains open, is similar to 
the Azbel-Hofstadter
problem~\cite{Azbel_But_nazo,Hofstadter_But_PRB76,Hasegawa_fluxS_PRB90} for the
weak magnetic field limit. It implies that the adiabatic invariant of the 
evolution can be given by the Chern number of the ground state
multiplet. This is correct as we discuss below.

\section{Adiabatic invariant}
As for the gapped ground state multiplet of anyons, we calculate the many-body 
Chern number~\cite{Niu_NTW_PRB85}
\begin{align}
 C=\frac{1}{2\pi i}\int_{T^2}d^2\eta F,
\end{align}
where $T^2=[0,2\pi]\times[0,2\pi]$, $F=(\pa A_y/\pa
\eta_x)-(\pa A_x/\pa \eta_y)$, $A_{x(y)}=\Tr[\Phi^\dagger(\pa \Phi/\pa 
\eta_{x(y)})]$ and 
$\Phi=\left(\ket{G_1}\cdots,\ket{G_{N_D}}\right)$ is a ground state multiplet. 
In the numerical calculation, we use the method proposed in Ref.~\onlinecite{
Fukui_FHS_JPSJ06}. 

\begin{figure}[t!]
  \begin{center}  
   \includegraphics[width=\columnwidth]{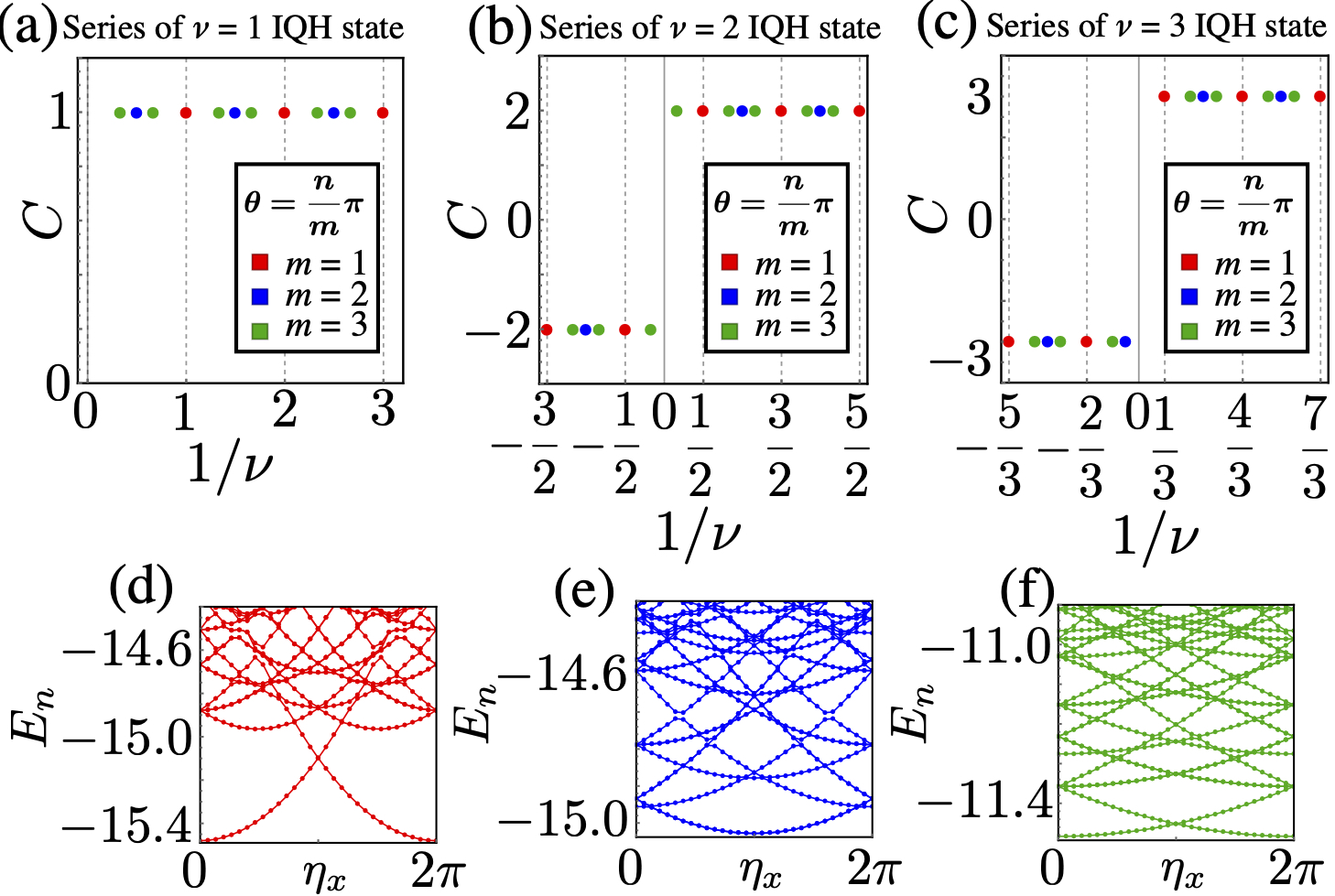}
  \end{center}
 \caption{
 (a)-(c) Chern number $C$ of the degenerated ground state multiplet are shown
 as functions of $1/\nu$. In (a), (b) and (c), we consider the systems in the 
 same setting as Fig.~\ref{fig:ah}(a), (b) and (c), respectively. 
 (d)-(f) Spectral flows at $1/\nu=0$ for each series. We set $\eta_y=0$. The 
 statistics parameter is given by (d) $\theta/\pi=2$, (e) $\theta/\pi=3/2$ and 
 (f) $\theta/\pi=4/3$, respectively. 
 }
 \label{fig:Chern}
\end{figure}
In Figs~\ref{fig:Chern}(a-c), we plot $C$ for the systems of the 
same setting as Figs.~\ref{fig:ah}(a-c). Although the dimensions of the 
multiplet changes wildly, the Chern number $C$ remains the same.
It suggests that $C$ is an adiabatic invariant of the 
evolution. As for a series that includes the $\nu=p$ IQH state, we numerically 
obtain
\begin{align}
 C=\text{sgn}(\nu)\times p,
 \label{eq:Chern}
\end{align}
where $\text{sgn}(x)$ is the sign function. For $\nu>0$, Eq.~\eqref{eq:Chern} 
is natural since the $\nu=p$ IQH state is included. However, the case
of $\nu<0$ is non-trivial since it does not include any simple state. 

While the energy of the QH systems is almost independent of $\vec{\eta}$,
the spectral flows at $1/\nu=0$ exhibit the strong $\vec{\eta}$ dependences,
see Figs.~\ref{fig:Chern}(d-f). They indicate the absence of the energy gap at 
$1/\nu=0$, which implies the Nambu-Goldston modes of the anyon superconductors.

\section{Topological degeneracy}
As mentioned above, the ground state degeneracy changes wildly during the 
evolution as shown in Figs.~\ref{fig:ah}(d-f). The fermion FQH state
at $\nu=p/q$ is $q$-fold degenerated~\cite{Haldane_TD_PRL85} but this pattern 
does not hold in the anyonic systems; the QH state with 
$(\nu,\theta/\pi)=(1,1/2)$ in Fig.~\ref{fig:ah}(e), for example, has 
$4$-fold degeneracy. We address this issue analytically below.

Let us consider a continuous translational invariant system of the size 
$L_x\times L_y$ with external magnetic field $B=\phi_0N_\phi/(L_xL_y)$. 
The results obtained below is valid even for lattice models as long as $\phi$ 
is sufficiently small, i.e., the magnetic length becomes much larger than the 
lattice constant.
The statistics of anyons is set $\theta=(n/m)\pi$ and a translation operator of
center-of-mass is given by 
$T(\bm{a})=\exp\{{(i/\hbar)\sum_i\bm{K}_i\cdot\bm{a}}\}$, where 
$\bm{K}_i=\bm{p}_i+e\bm{A}(\bm{r}_i)-eB\bm{e}_z\times\bm{r}_i$~\cite{Zak_MagTran_PF64,Haldane_TD_PRL85,Tao_TD_PRB86}.
Since the interactions of the system including the statistical vector 
potential~\cite{Wilczek_anyon2_PRL82} are given by
the relative coordinates of anyons, $T(\bm{a})$ commutes with the 
Hamiltonian $H$. Noting that
$T(\bm{b})^{-1}T(\bm{a})^{-1}T(\bm{b})T(\bm{a})
=e^{i\frac{e}{\hbar}B(\bm{a}\times\bm{b})N_a}$,
let us now assume the followings
\begin{align}
 &\rho_i^{-1}T(\bm{a})^{-1}\rho_iT(\bm{a})
 =e^{i\frac{e}{\hbar}B(\bm{a}\times L_y\bm{e}_y)}
 \label{eq:rhoT}\\
 &T(\bm{b})^{-1}\tau_i^{-1}T(\bm{b})\tau_i
 =e^{i\frac{e}{\hbar}B(L_x\bm{e}_x\times\bm{b})},
 \label{eq:tauT}
\end{align}
since each loop given by Eqs.~\eqref{eq:rhoT} and
~\eqref{eq:tauT} does not enclose the other anyons.

Equation~\eqref{eq:braid} implies $[\tau_i^m,\rho_j]=0$. (The proof for any 
$i$ and $j$ is given in Appendix~\ref{sec:braid}). Then by defining 
$\mathcal{T}_A\equiv T(\frac{1}{m}\frac{L_y}{N_\phi}\bm{e}_y)$, which 
satisfies
\begin{align}
 [\mathcal{T}_A,\tau_i^m]=[\mathcal{T}_A,\rho_i]=0,
\end{align}
let us take the simultaneous eigenstate $\ket{\psi_0}$, which satisfies
$H(\vec{\eta})\ket{\psi_0}=E(\vec{\eta})\ket{\psi_0}$ and 
$\mathcal{T}_A\ket{\psi_0}=e^{i\lambda}\ket{\psi_0}$ with $\lambda$ real.
Here, the twisted boundary angles $\vec{\eta}$ are specified by 
$\tau_i^m$ and $\rho_i$ with
Eqs.~\eqref{eq:Wx} and ~\eqref{eq:Wy}. Further defining 
$\mathcal{T}_B\equiv T(\frac{L_x}{N_\phi}\bm{e}_x)$ and 
$\mathcal{T}_C\equiv \tau_1T(\frac{n}{m}\frac{L_x}{N_\phi}\bm{e}_x)$, we 
define a new state 
$\ket{\psi_{s,t}}\equiv\mathcal{T}_B^s\mathcal{T}_C^t\,\ket{\psi_0}$.
While $\mathcal{T}_B$ and $\mathcal{T}_C$ commute with 
$\tau^m_i$ and $\rho_i$, we have
\begin{align}
 &\mathcal{T}_A\mathcal{T}_B=
 \mathcal{T}_B\mathcal{T}_A
 e^{i2\pi\frac{1}{m}\nu},\\
 &\mathcal{T}_A\mathcal{T}_C=
 \mathcal{T}_C\mathcal{T}_A
 e^{i2\pi(\frac{1}{m}+\nu\frac{n}{m^2})}.
 \label{eq:ACCA}
\end{align}
It implies $H(\vec{\eta})\ket{\psi_{s,t}}=E(\vec{\eta})\ket{\psi_{s,t}}$ and 
$\mathcal{T}_A\ket{\psi_{s,t}}=e^{i\lambda}e^{i2\pi f_{s,t}}
\ket{\psi_{s,t}}$ with
\begin{align}
 f_{s,t}=\frac{\nu s+\left(1+\nu\theta/\pi\right)t}{m}
 =\frac{\nu}{pm}\left(p\left(s+t\right)+t\right),
\end{align}
where $\nu=p/[p(1-\theta/\pi)+1]$ is used at the last part. Thus, the 
topological 
degeneracy $N_\text{TD}$ is given by the number of pairs $(s,t)$ that give
different values of $f_{s,t}\mod1$. 
Since $pm/\nu$ is always integer, one obtains $N_\text{TD}=pm/|\nu|$.
Using Eq.~\eqref{eq:Chern} and $M=m$ (irreducible representation), we have
\begin{align}
 N_\text{TD}=MC/\nu.
 \label{eq:TD}
\end{align}
This is consistent with the obtained ground state degeneracy shown in 
Fig.~\ref{fig:ah}(d-f).
Anyon nature shown in Eq.~\eqref{eq:ACCA} gives the extra
degeneracy compared with the fermionic standard case~\cite{Haldane_TD_PRL85}.

\section{Generalized Streda formula}
Taking difference of Eq.~\eqref{eq:TD} for two possible cases in a series,
one obtains $\Delta N_\text{TD}/\Delta(M/\nu)=C$, where we assume the Chern 
number $C$ is the invariant. Since $M/\nu=MN_xN_y\phi/N_a$ with $\phi$
the number of the flux per plaquette, we finally have 
\begin{align}
 \frac{\Delta(N_p/N'_\text{site})}{\Delta\phi}=C,
 \label{eq:streda}
\end{align}
where $N_p\equiv N_\text{TD}N_a$ is the ``parton'' number corrected by the 
topological degeneracy and $N'_\text{site}\equiv MN_xN_y$ is the extended 
number of sites due to the non-Abelian nature of the representation. This is a 
{\it generalized} Streda formula for anyons. Note that Eq.~\eqref{eq:streda} 
for fermions ($M=1$, $\nu=p/q$, $N_\text{TD}=q$, $C=p$) reduces to the standard
Streda formula~\cite{Streda_StreFor_IOP82}. 
When one includes a reducible representation of the braid group, i.e., $M=tm$ 
with $2\leq t$ ($t$: integer), the degeneracy $N_\text{TD}$ increases by $t$ 
times. Therefore, Eqs.~\eqref{eq:TD} and \eqref{eq:streda} holds generally.

\section{Conclusion}
In this Letter, the adiabatic heuristic principle for the QH states is 
demonstrated on a torus numerically. The 
emergence of the anyon superconducting states is also suggested. The 
Chern number of the ground state multiplet serves as the adiabatic 
invariant of the evolution although their degeneracy changes 
wildly. The anyon nature brings the extra multiplicity to the topological 
degeneracy. It results in a generalized Streda formula that follows from the 
translational invariance. Extensions of this adiabatic principle on a 
torus can be useful to characterize the non-Abelian FQH states.
\\

\begin{acknowledgments}
 We thank the Supercomputer Center, the Institute for Solid State
 Physics, the University of Tokyo for the use of the facilities.
 The work is supported in part by JSPS KAKENHI Grant Numbers JP17H06138 
 (K.K, Y.H.), and JP19J12317 (K.K.).
\end{acknowledgments}

\appendix
\section{Constraints on statistical phase}
\label{sec:braid}
In this appendix, we derive the relation $1/\nu+\theta/\pi=\text{const.}$ from 
the braid group analysis on a torus. Also, a proof of $[\tau_i^m,\rho_j]=0$ 
for any $i$ and $j$ is also given here.

In the main text, we denote the generators of the braid group on a torus by 
$\sigma_i$, $\tau_i$ and $\rho_i$. They satisfy the following 
relations~\cite{Einarsson_Braid_PRL90,Einarsson_Braid_MPLB,Li_Braid_B93}:
\begin{align}
 &\tau_{i+1}^{-1}\rho_{i}\tau_{i+1}\rho_{i}^{-1}
 =(\sigma_i^{-1})^2,
 \label{eq:tauirhoj}\\
 &\rho_1^{-1}\tau_1^{-1}\rho_1\tau_1
 =\sigma_1\cdots\sigma_{N_a-1}
 \sigma_{N_a-1}\cdots\sigma_1e^{i\frac{e}{\hbar}BL_xL_y},
 \label{eq:tauirhoi}\\
 &\tau_{i+1}
 =\sigma_i^{-1}\tau_i\sigma_i^{-1}e^{i\frac{e}{\hbar}
 (\alpha_{i+1}-\alpha_i)},
 \label{eq:tautau}\\
 &\rho_{i+1}
 =\sigma_i\rho_i\sigma_ie^{i\frac{e}{\hbar}
 (\beta_{i+1}-\beta_i)},
 \label{eq:rhorho}
\end{align}
where $\alpha_i$ and $\beta_i$ are real numbers. 
Equation~\eqref{eq:tauirhoj} is the same as Eq.~\eqref{eq:braid}. 
The derivations of Eqs.~\eqref{eq:tautau} and ~\eqref{eq:rhorho} are 
given in Appendix~\ref{sec:noncontractible}.

Substituting $\sigma_i=e^{i\theta}\bm{1}_M$ into 
Eqs.~\eqref{eq:tauirhoj}, \eqref{eq:tauirhoi}, \eqref{eq:tautau} and 
\eqref{eq:rhorho}, we have
\begin{align}
 &\rho_{i}\tau_{i+1}=\tau_{i+1}\rho_{i}e^{-i2\theta},
 \label{eq:braid1}\\
 &\rho_1\tau_1=\tau_1\rho_1e^{i2(N_a-1)\theta+i2\pi N_\phi},
 \label{eq:braid2}\\
 &\tau_{i+1}=\tau_ie^{-i2\theta}e^{i\frac{e}{\hbar}(\alpha_{i+1}-\alpha_i)},
 \label{eq:braid3}\\
 &\rho_{i+1}=\rho_ie^{i2\theta}e^{i\frac{e}{\hbar}(\beta_{i+1}-\beta_i)}.
 \label{eq:braid4}
\end{align}
Here, we note that the representations
\begin{align}
 \tau_j&=e^{i\frac{e}{\hbar}\alpha_j}e^{-i2\theta(j-1)}
 W_x,
  \label{eq:taumatrix}\\
 \rho_j&=e^{i\frac{e}{\hbar}\beta_j}e^{i2\theta(j-1)}
 W_y,
 \label{eq:rhomatrix}
\end{align}
satisfy the relations in Eqs.~\eqref{eq:braid1}, \eqref{eq:braid2}, 
\eqref{eq:braid3} and \eqref{eq:braid4}.
Substituting Eq.~\eqref{eq:braid3} into Eq.~\eqref{eq:braid1}, one gets
\begin{align}
 &\rho_{i}\tau_{j}=\tau_{j}\rho_{i}e^{-i2\theta}.
 \label{eq:braid5}
\end{align}
If $\theta/\pi=n/m$, it reduces to $[\tau_i^m,\rho_j]=0$. Comparing 
Eq.~\eqref{eq:braid5} 
for $i=j=1$ with Eq.~\eqref{eq:braid2}, we get 
\begin{align}
 e^{i2\theta N_a+i2\pi N_\phi}=1.
 \label{eq:const}
\end{align}
It implies that $\theta/\pi+1/\nu=2\pi s/N_a$ with $s$ integer.

\begin{figure}[t!]
  \begin{center}  
   \includegraphics[width=\columnwidth]{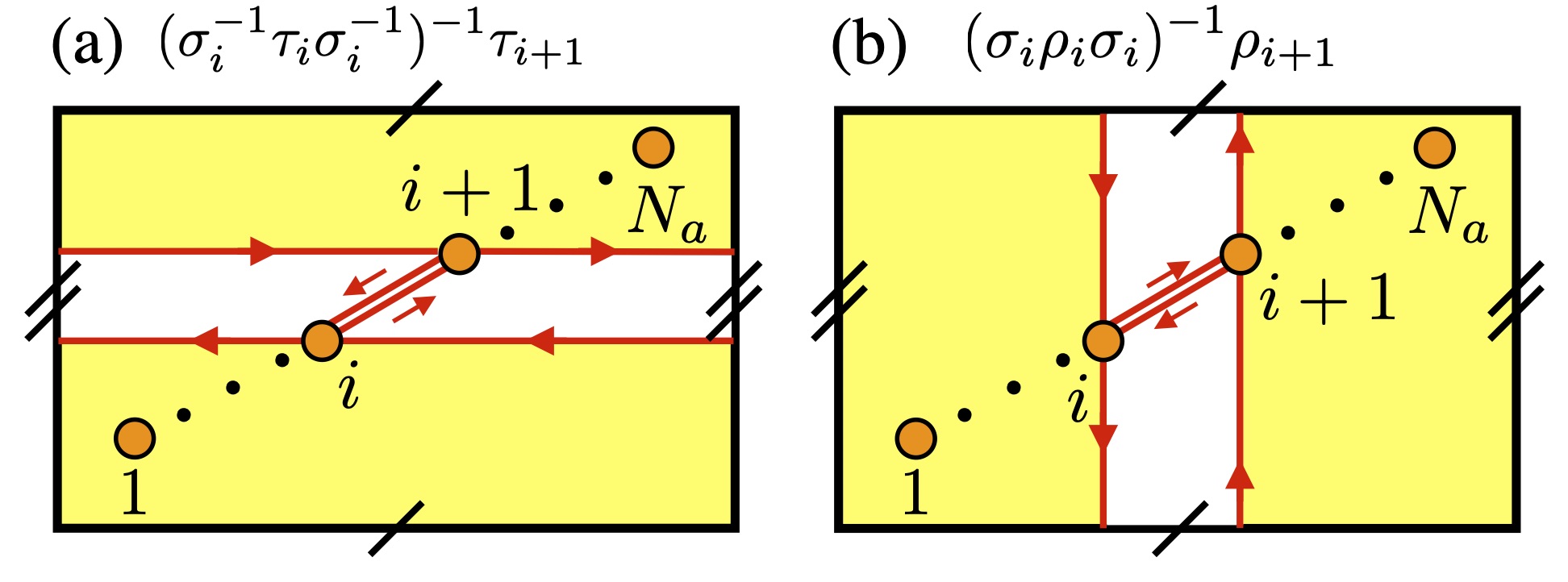}
  \end{center}
 \caption{
 Closed paths defined by (a) 
 $(\sigma_i^{-1}\tau_i\sigma_i^{-1})^{-1}
 \tau_{i+1}$ and 
 $(\sigma_i\rho_i\sigma_i)^{-1}\rho_{i+1}$.
 }
 \label{fig:taurho}
\end{figure}
\section{Noncontractible loops on a torus}
\label{sec:noncontractible}
In this appendix, we prove Eqs.~\eqref{eq:tautau} and \eqref{eq:rhorho}. If 
the magnetic flux is absent, the relations of the braid group
are given as~\cite{Birman_Braid_M69,Einarsson_Braid_PRL90}
\begin{align}
 \tau_{i+1}
 &=\sigma_i^{-1}\tau_i\sigma_i^{-1},\\
 \rho_{i+1}
 &=\sigma_i\rho_i\sigma_i.
\end{align}
Note that $(\sigma_i^{-1}\tau_i\sigma_i^{-1})^{-1}\tau_{i+1}$  and 
$(\sigma_i\rho_i\sigma_i)^{-1}\rho_{i+1}$ move anyons along closed loops shown 
in Figs.~\ref{fig:taurho}(a) and (b), respectively. Therefore, if the magnetic 
field described by the vector potential $\bm{A}(\bm{r})$ is present, 
$(\alpha_{i+1}-\alpha_i)/\phi_0$ and $(\beta_{i+1}-\beta_i)/\phi_0$ fluxes
penetrate each closed paths, respectively, where
$\alpha_i=\oint_{L_{\tau_i}}d\bm{r}\cdot\bm{A}(\bm{r})$,
$\beta_i=\oint_{L_{\rho_i}}d\bm{r}\cdot\bm{A}(\bm{r})$, and 
$L_{\tau_i(\rho_i)}$ is the path given by $\tau_i(\rho_i)$. Then we obtain
Eqs.~\eqref{eq:tautau} and \eqref{eq:rhorho}.

\bibliographystyle{apsrev4-1}
\bibliography{citation}

\begin{thebibliography}{47}
\bibitem [{tor()}]{torus}%
  \BibitemOpen
  \href@noop {} {}\bibinfo {note} {Since the system of anyons is intrinsically
  a many-body problem, the construction of the pseudopotential projected into
  the lowest Landau level is impossible. Therefore, we choose a toroidal
  lattice model for the direct numerical demonstrations. This geometry is also
  suitable for evaluation of the topological numbers.}\BibitemShut {Stop}%
\end{thebibliography}

\end{document}